\documentclass[12pt]{article}
\usepackage{amsmath}
\usepackage{graphicx}
\usepackage{float}
\usepackage{enumerate}
\usepackage{url} 
\usepackage{amsfonts,amsmath,verbatim,amssymb,epsfig,enumitem,amsbsy,enumitem,color,graphicx}
\usepackage{algorithm, algorithmic}
\usepackage{bm,ulem,courier}
\usepackage{subfigure}
\usepackage{float}
\usepackage{multirow}
\usepackage{diagbox}
\usepackage[figuresright]{rotating}
\usepackage{pdflscape}
\usepackage{booktabs}
\usepackage{bm}
\usepackage{makecell}
\usepackage{array}
\usepackage{amsthm}
\usepackage{makecell}
\usepackage{natbib}
\usepackage{xcolor}
\usepackage{hyperref}

\setcitestyle{maxcitenames=3, mincitenames=1}

\newcommand{\blind}{1}

  \renewenvironment{thebibliography}[1]{
    \begin{oldthebibliography}{#1}
      \setlength{\parskip}{0ex}
      \setlength{\itemsep}{0.ex}}
  {\end{oldthebibliography}}

\newcommand{\myvec}[1]%
{\stackrel{\raisebox{-2pt}[0pt][0pt]{\tiny$\rightharpoonup$}}{#1}}

\long\def\red#1{{\color{black}#1}}

\makeatletter
\newenvironment{breakablealgorithm}
  {
   \begin{center}
     \refstepcounter{algorithm}
     \hrule height.8pt depth0pt \kern2pt
     \renewcommand{\caption}[2][\relax]{
       {\raggedright\textbf{\ALG@name~\thealgorithm} ##2\par}%
       \ifx\relax##1\relax 
         \addcontentsline{loa}{algorithm}{\protect\numberline{\thealgorithm}##2}%
       \else 
         \addcontentsline{loa}{algorithm}{\protect\numberline{\thealgorithm}##1}%
       \fi
       \kern2pt\hrule\kern2pt
     }
  }{
     \kern2pt\hrule\relax
   \end{center}
  }
\makeatother

\makeatletter
\newcommand*{\centerfloat}{%
  \parindent \z@
  \leftskip \z@ \@plus 1fil \@minus \marginparwidth
  \rightskip \leftskip
  \parfillskip \z@skip}
\makeatother

\newtheorem{thm}{Theorem}

\newtheorem{prop}{Proposition}

\newtheorem{assump}{Assumption}
\newtheorem{remark}{Remark}

\DeclareMathOperator*{\argmin}{arg\,min}
\DeclareMathOperator*{\argmax}{arg\,max}

\begin{document}

\def\spacingset#1{\renewcommand{\baselinestretch}%
{#1}\small\normalsize} \spacingset{1}


\if1\blind
{
  \title{\bf  Deconfounding via Profiled Transfer Learning}
 \author{
    Ziyuan Chen$^{1}$\footnotemark[2],
    Yifan Jiang$^{2}$\footnotemark[2],
    Jingyuan Liu$^{3}$\footnotemark[1],
    Fang Yao$^{1}$ \\
    \\
    $^{1}$School of Mathematical Sciences, Center for Statistical Science,\\ Peking University, Beijing, China \\
        $^{2}$Department of Statistics, Eberly College of Science,\\The Pennsylvania State University, State College, PA, United States\\
    $^{3}$MOE Key Laboratory of Econometrics,\\ Department of Statistics and Data Science in School of Economics, \\Laboratory of Digital Finance, 
Xiamen University, Xiamen, China 
}
\date{}
\maketitle
\footnotetext[2]{Ziyuan Chen and Yifan Jiang are co-first authors.}
\footnotetext[1]{Jingyuan Liu is the corresponding author, E-mail: \texttt{jingyuan@xmu.edu.cn}. \\
This research is partially supported by the National Key Research and Development Program of China (No.2022YFA1003800), the National Natural Science Foundation of China (No. 12292981, 12526213, 12271456, 71988101, 123B2009), 
the Fundamental and Interdisciplinary Disciplines Breakthrough Plan of the Ministry of Education of China (No. JYB2025XDXM118), 
the Ministry of Education Research in the Humanities and Social Sciences (No. 22YJA910002), 
the New Cornerstone Science Foundation, the LMAM, 
the Fundamental Research Funds for the Central Universities, Peking University, and LMEQF.}
} \fi

\if0\blind
{
  \bigskip
  \bigskip
  \bigskip
  \begin{center}
    {\LARGE\bf Deconfounding via Profiled Transfer Learning}
\end{center}
\date{}
\medskip
} \fi

\bigskip
\begin{abstract}
Unmeasured confounders are a major source of bias in regression-based effect estimation and causal inference. In this paper, we propose a new profiled transfer learning framework, ProTrans, to address confounding effects in the target dataset, when additional source datasets with similar confounding structures are available. We introduce the concept of profiled residuals to characterize the shared confounding patterns between source and target datasets. By incorporating these profiled residuals into the target debiasing step, we effectively mitigate the latent confounding effects. We also propose a source selection strategy to enhance the robustness of ProTrans to noninformative sources. As a byproduct, ProTrans can also be used to estimate treatment effects in the presence of potential confounders, without the use of auxiliary features such as instrumental or proxy variables, which are often challenging to select in practice. Theoretically, we prove that the resulting estimated model shift from the sources to the target is confounding-free without imposing specific assumptions on the true confounding structure, and that the target parameter estimation achieves the minimax optimal rate under mild conditions. Simulated and real-world experiments validate the effectiveness of ProTrans and support the theoretical findings.
\end{abstract}

\noindent%
{\it Keywords: Confounded models; Transfer learning; Deconfounding; Model shift} 
\vfill

\newpage
\spacingset{1.9} 
\section{Introduction}

In many real-world applications ranging from genetics to economics, unmeasured confounders frequently arise. Failure to account for these confounders can lead to biased parameter estimation, spurious correlations, misleading partial dependencies and invalid causal conclusions.
Existing methods for addressing hidden confounders typically involve either direct estimation of confounding effects or the incorporation of auxiliary variables. For example, factor analysis-based techniques attempt to model the confounding structure directly \citep{cevid2020spectral,fan2021robust,guo2022doubly,bing2022adaptive,sun2024decorrelating}.  Instrumental variable (IV) techniques address unmeasured confounding by leveraging exogenous variables that affect the outcome only through the exposure of interest \citep{baiocchi2014instrumental, maciejewski2019using}.
\cite{miao2018identifying} and \cite{cui2024semiparametric} use observed proxy variables for hidden confounders to adjust for them in the models, and difference-in-differences methods \citep{donald2007inference,ding2019bracketing} can mitigate time-invariant hidden confounding. Nevertheless, these approaches often face practical challenges in choosing suitable auxiliaries or rely on stringent assumptions, such as the exclusion restriction and the parallel trends assumption \citep{martens2006instrumental,zhao2025semiparametric}. 

To circumvent these issues, it is important to note that the aforementioned traditional deconfounding methods operate within single-dataset frameworks. However, modern data collection often provides access to multiple data sources that may share similar or related data structures with the target dataset of primary interest. Transfer learning is therefore used to leverage source information and improve estimation and inference performance when modeling the target dataset. Although related, source models often differ from the target model to a certain extent. Consequently, a growing body of work focuses on modeling and addressing distributional differences and heterogeneity between source and target models; see, for example, \cite{li2022transfer, lin2024profiled, tian2025learning, yuan2025optimal}. For a broad and in-depth understanding of transfer learning, see comprehensive surveys such as \cite{weiss2016survey,gholizade2025review}. The similarity between source and target datasets reasonably suggests that their underlying confounding mechanisms may also exhibit comparable patterns. This motivates us to leverage source information to mitigate confounding bias in the target data through transfer learning.

From the statistical perspective, existing transfer learning methodologies generally follow two paradigms: (1) a two-step approach that first estimates an ensemble model incorporating all source datasets, followed by target-specific model shift estimation \citep{li2022transfer,tian2023transfer,cai2024transferb}, and (2) a joint estimation framework that simultaneously learns target and source parameters \citep{he2024adatrans,li2024estimation}. 
However, direct application of either paradigm to confounded models fails to exploit the shared confounding structure between the target and sources. We will later provide a detailed demonstration of this defect in conventional transfer learning methods. As a result, the estimates for model shift and consequently the target parameters remain biased. 


To this end, we propose a new profiled transfer learning framework, referred to as ProTrans, for scenarios in which the source and target models share similar hidden confounding structures. Beyond transferring covariates-response associations from the source data, ProTrans also transfers hidden confounding effects from sources to deconfound the target model. This is achieved by constructing source profiled residuals and transferring them to the target. The source profiled residuals are defined as the differences between observed and predicted values from the initially estimated source models, which preserve the confounding structures. By transferring these residuals to the target, ProTrans effectively aligns the confounding effects between sources and the target. By subtracting the transferred profiled residuals during the target-specific debiasing step, ProTrans eliminates confounding bias in the estimation of model shift. The target parameters can be naturally estimated using the resulting model shift and certain deconfounded source parameter estimates. Furthermore, under more general and realistic scenarios where confounding effects may vary between several sources and the target model, we introduce an informative source selection procedure based on a newly proposed quality score for each source.  

We systematically study the theoretical properties of ProTrans estimators for both model shift and the target parameter. Noting that model shift can be particularly important in certain applications, we show that the ProTrans estimator for model shift achieves the oracle convergence rate, matching the optimal estimation rate of models without hidden confounders. Unlike existing methods that often rely on strict assumptions about confounders, ProTrans does not require any specific assumptions regarding hidden confounders, making it applicable to a much wider range of scenarios. For target parameter estimation, we show that the hidden confounders only affect the estimation of source models while this influence can be substantially mitigated due to the large size of the source data.
As a result, the estimation of the target model in ProTrans is faster than that achieved by classical transfer learning methods, in which the confounding effect on the target model is still difficult to eliminate due to the small size of the target data and hence tends to dominate the resulting estimation error. We also provide the minimax lower bound for the target model estimation under mild conditions. Moreover, with the newly proposed source selection procedure, we show that ProTrans is robust to low-quality sources, and the selection procedure can effectively filter out unhelpful sources while preserving the theoretical guarantees. 

The contributions of ProTrans can be summarized in two main aspects. First, to the best of our knowledge, ProTrans is the first attempt to leverage additional source information to mitigate confounding effects. 
Notably, ProTrans requires neither restrictive assumptions about hidden confounding structures nor auxiliary variables as in most existing deconfounding methods, significantly expanding its applicability to real-world problems. Second, ProTrans also complements existing transfer learning literature, since direct application of conventional transfer learning to confounded models fails to transfer the confounding structure and hence cannot eliminate confounding effects. As a byproduct, ProTrans holds particular interest for treatment effect estimation, where the treatment group serves as the target and the typically larger control group serves as the source. In this framework, the model shift directly yields treatment effect estimates.


The remainder of the paper is organized as follows. Section \ref{sec:methodology} introduces the ProTrans method for estimating both the model shift and the target parameter, and demonstrates that conventional transfer learning approaches are biased in the presence of confounding. Section \ref{sec:theorem} establishes the non-asymptotic bounds and minimax optimality of ProTrans estimators for model shift and the target parameter. Section \ref{sec:source_selection} introduces a source selection procedure to handle noninformative source data. Sections \ref{sec:simulation} and \ref{sec:realdata} evaluate the finite-sample performance and demonstrate the application of ProTrans via simulation studies and real-data applications, respectively. Additional theoretical results, technical proofs and numerical results are provided in the online Supplementary Material.

\section{Deconfounding via Profiled Transfer Learning}
\label{sec:methodology}
\subsection{Confounded model: challenge and motivation}
\label{subsec:preliminary}

Consider the confounded model with data $(X_{ti}, Y_{ti})$ drawn from our target population:
\begin{align}
\label{eq:target_model}
    Y_{ti} = \beta_t^\top X_{ti} + \phi_t^\top H_{ti} + \varepsilon_{ti},\quad X_{ti} = f(H_{ti}) +E_{ti}, \ i = 1, \dots, n_t,
\end{align}
where $n_t$ is the target sample size, which is typically small. The coefficient vector $\beta_t\in \mathbb{R}^p$, which links the response variable $Y_{ti} \in \mathbb{R}$ to the observed covariates $X_{ti} \in \mathbb{R}^p$, is often of primary interest. Let $H_{ti}\in \mathbb{R}^q$, where $q$ is fixed, denote the hidden confounders, and $\phi_t^\top H_{ti}$, where $\phi_t\in \mathbb{R}^q$, represents the confounding effect in the target model. The function $f: \mathbb{R}^{q}\to\mathbb{R}^p$ is an unknown mapping that describes the confounding structure in $X_{ti}$ and the error term $E_{ti}\in \mathbb{R}^p$ is assumed to be uncorrelated with $H_{ti}$. 
The pure noise term $\varepsilon_{ti}\in \mathbb{R}$ is assumed to be independent of $X_{ti}, \ H_{ti}$ and $E_{ti}$. For identification, we further assume that $X_{ti},Y_{ti},H_{ti},E_{ti}$ and $\varepsilon_{ti}$ all have zero mean. 

Model \eqref{eq:target_model} is a structural equation model, which is widely used to describe the general causal relationships among covariates, response, and hidden confounders \citep{fox2002structural,pearl2012causal}. The presence of hidden confounders $H_{ti}$ inevitably introduces bias when estimating $\beta_t$. Denote the design matrix for the covariates by $X_t=(X_{t1},\ldots,X_{tn_t})^\top\in\mathrm{R}^{n_t\times p}$, and analogously denote $Y_t\in \mathbb{R}^{n_t}, H_t\in \mathbb{R}^{n_t\times q},E_t\in \mathbb{R}^{n_t\times p}$ and $\varepsilon_t\in \mathbb{R}^{n_t}$.
Specifically, the bias can be identified by rewriting the model as 
\begin{align}
    Y_{t} =  X_{t}(\beta_t+b_t) + U_{t}+\varepsilon_{t},
    \label{perturb}
\end{align}
where the bias $b_t =(X_t^\top X_t)^{\dagger}X_t^\top H_t \phi_t$, $A^{\dagger}$ denotes the Moore–Penrose inverse of $A$, and $U_{t} = (I_{n_t} - X_t (X_t^\top X_t)^{\dagger}X_t^\top) H_t\phi_t$ is an term uncorrelated with $X_t$.

In the high-dimensional setting, \cite{cevid2020spectral,guo2022doubly,sun2024decorrelating} proposed using the trim transform approach to mitigate this bias when $f(\cdot)$ is linear. Define the singular value decomposition of $X_t$, $X_t = U_tD_tV^\top_t$. Then the trim transform suggests to take
\begin{align}
\label{eq:trim_transform}
    \mathcal{Q}_t = U_t \operatorname{diag}\left(\min(\tau_t/D_t,1)\right)U_t^\top,
\end{align}
where $\tau_t$ is the $\tau$-th ordered singular value of $X_t$ for some $\tau$. Finally, $\beta_t$ can be estimated from the trimmed response $\tilde{Y}_t = \mathcal{Q}_t Y_t$ and trimmed covariates $\tilde{X}_t=\mathcal{Q}_t X_t$ using standard regularized regression methods such as LASSO.
The effectiveness of this approach heavily relies on the concentration of the bias $b_t$ in the spike directions of the empirical covariance matrix $X_t^\top X_t/n_t$. The trim transform needs to shrink the original data in directions such that the bias $b_t$ is much smaller than the signal $\beta_t$. This imposes a strict limitation on the method, restricting its use to the model \eqref{eq:target_model} with a linear form $f(\cdot)$. A similar limitation arises under the fixed $p$ scenario, as \cite{tang2023synthetic} shows that an adequately constructed synthetic IV must also be projected out.

Instead of deconfounding the model \eqref{eq:target_model} using target data alone, we often have access to source datasets that share similar confounding structures with the target. For example, in genetic studies, hidden confounders such as cell-type composition and batch effects are often shared across datasets. Data from historical records or other tissues with large sample sizes can therefore be leveraged. Thus a natural solution is to utilize the rich source data via transfer learning where a typical challenge is to estimate the model shift from the source to the target model \eqref{eq:target_model}. Interestingly, similar confounding structures help achieve confounding-free model-shift estimation, which we view as a ``blessing of confounding''.  This can be useful in certain causal inference problems with hidden confounders. Furthermore, deconfounding the source models is more feasible due to their larger sample sizes. This allows for sufficiently precise deconfounding estimation even when the linear constraint on $f(\cdot)$ is relaxed to broader, unknown functional forms. Leveraging source information helps to handle more general confounder models and improves estimation relative to relying solely on target data.


\subsection{ProTrans: Profiled transfer learning}

Assume $K$ source datasets are available. For each $k, \ k=1,\ldots,K$, we observe i.i.d. source data $(X^{(k)}_{s},Y^{(k)}_{s})$ from the $k$-th source model 
\begin{align}
\label{eq:source_model}
    Y^{(k)}_{si} = \beta^{(k)\top}_s X^{(k)}_{si} + \phi^{(k)\top}_s H^{(k)}_{si} + \varepsilon^{(k)}_{si},\quad X^{(k)}_{si} = f(H^{(k)}_{si}) +E^{(k)}_{si}, \ i=1,\cdots,n_s^{(k)},
\end{align}
where the notations are parallel to those in the target model \eqref{eq:target_model}.  
Denote the total sample size of all sources as $n_s = \sum_{k=1}^K n_s^{(k)}$, and assume $n_s\gg n_t$.  
In order to utilize the extensive source information to improve the estimation of target parameters, transfer learning frameworks typically decompose the parameter of interest $\beta_t$ in the target model \eqref{eq:target_model} into two components: the average source effect $\beta_s = \sum_{k=1}^K n_s^{(k)}\beta_s^{(k)}/n_s$ and the model shift $\eta_0 = \beta_t-\beta_s$ from sources to target. 

We first focus on the model shift $\eta_0$, and propose a profiled transfer technique that leverages the shared confounding structure between the source models \eqref{eq:source_model} and the target model \eqref{eq:target_model} to mitigate the estimation bias of $\eta_0$. 
Let $\hat{\beta}_{\text{init}}$ be some initial estimator for $\beta_s$. Note that $\hat{\beta}_{\text{init}}$ requires only mild conditions (to be stated in Theorem \ref{thm:upper_bound_of_eta}), and need not be deconfounded in advance. Then we define the profiled source residual $\hat{Z}_s^{(k)} = Y^{(k)}_s - X^{(k)}_s\hat{\beta}_{\text{init}}\in \mathbb{R}^{n_s^{(k)}}$. Intuitively, the residual $\hat{Z}_s^{(k)}$ encodes information about hidden confounders in the $k$-th source model, specifically,
\begin{align*}
    \hat{Z}_s^{(k)} = X_s^{(k)}\left(\beta_s^{(k)} - \hat{\beta}_{\text{init}}\right) + H_s^{(k)} \phi_s^{(k)} + \varepsilon_s^{(k)}.
\end{align*}

We then propose a profiling strategy to transfer the confounding effects $H_s\phi_s^{(k)}$ from sources to the target using $\hat{Z}_s^{(k)}$, by constructing a profiled target residual $\hat{Z}_t\in \mathbb{R}^{n_t}$ that mitigates the confounding effect in the target data. The goal of $\hat{Z}_t$ is to counteract the confounding effect in the target model so that $(Y_t - \hat{Z}_t)$ is minimally influenced by $H_t$. In the target-specific debiasing step, we subtract both $\hat{Z}_t$ and $X_t\hat{\beta}_{\text{init}}$ from the target response and then apply classical (penalized) least-squares estimators. For example, a confounding-free LASSO estimator for the model shift $\eta_0 = \beta_t - \beta_s$ can be obtained as
\begin{align}
\label{eq:estimator_of_eta}
    \hat{\eta} = \argmin_{\eta \in \mathbb{R}^p }\left\{\frac{1}{n_t}\left\|Y_t - X_t \hat{\beta}_{\text{init}} - \hat{Z}_t - X_t \eta \right\|_2^2 +  \lambda_t \|\eta\|_1\right\},
\end{align}
under the high-dimensional setting with a sparse $\eta_0$. 

To construct a reasonable profiled target residual $\hat{Z}_t$, note that $Y_t -\hat{Z}_t - X_t\hat{\beta}_{\text{init}}$ in \eqref{eq:estimator_of_eta} can be decomposed as
\begin{align*}
    Y_t - \hat{Z}_t - X_t\hat{\beta}_{\text{init}}&= X_t\left(\eta_0 +(X_t^\top X_t)^{\dagger} X_t^\top (H_t\phi_t - \hat{Z}_t) - (\hat{\beta}_{\text{init}} - \beta_s)\right) \\&+ \left(I_{n_t} - X_t(X_t^\top X_t)^{\dagger} X_t^\top\right) \left(H_t\phi_t -\hat{Z}_t\right) + \varepsilon_t,
\end{align*}
The role of $\hat{Z}_t$ is to ensure that the term $(X_t^\top X_t)^{\dagger} X_t^\top (H_t\phi_t - \hat{Z}_t)$ approximates the bias $\hat{\beta}_{\text{init}}-\beta_s$ induced by the initial estimator, while remaining unaffected by the confounders. More precisely, by further decomposing the least squares term $\|Y_t - \hat{Z}_t - X_t\hat{\beta}_{\text{init}}\|_2^2$, we should ensure that the additional error term
\begin{align}
\label{eq:sup:additional_error_term}
    & \left|\left(\eta_0 - \hat{\eta}\right)^\top X_t^\top X_t \left((X_t^\top X_t)^{\dagger} X_t^\top (H_t\phi_t - \hat{Z}_t) - \left(\hat{\beta}_{\text{init}} - \beta_s \right)\right)\right|\notag\\ \le\ &
    \left\|\eta_0 - \hat{\eta}\right\|_1 \left\| X_t^\top (H_t\phi_t - \hat{Z}_t) -X_t^\top X_t \left(\hat{\beta}_{\text{init}} - \beta_s \right)\right\|_{\infty}
\end{align}
is sufficiently small. When the source and target models have similar structures, we have $X_t^\top X_t/n_t \approx X_s^\top X_s/n_s$ and $X_t^\top H_t\phi_t/n_t \approx \sum_{k=1}^K X_s^{(k)\top}H_s^{(k)}\phi_s^{(k)}/n_s$ under the infinite norm. Recalling the definition of $\hat{Z}_s^{(k)}$, we obtain
\begin{align*}
    \frac{1}{n_s}\sum_{k=1}^K X_s^{(k)\top}\hat{Z}_s^{(k)} \approx  \frac{1}{n_t} X_t^\top H_t\phi_t - \frac{1}{n_t}X_t^\top X_t\left( \hat{\beta}_{\text{init}}-\beta_s \right) +  \frac{1}{n_s}\sum_{k=1}^K X_s^{(k)\top}\varepsilon_s^{(k)}.
\end{align*}
Therefore, it is sufficient to align $X_t^\top \hat{Z}_t$ with $X_s^{(k)\top}\hat{Z}_s^{(k)}$ to guarantee that the additional error term \eqref{eq:sup:additional_error_term} remains small. Thus, we construct the profiled target residual $\hat{Z}_t$ through the profiled residual transfer:
\begin{align}
\label{eq:bias_correction_term}
    \hat{Z}_t = \argmin_{Z_t \in \mathbb{R}^{n_t}}\left\{\left\|\frac{1}{n_t}X_t^\top Z_t - \frac{1}{n_s}\sum_{k=1}^K X_s^{(k)\top}\hat{Z}_s^{(k)}\right\|_{\infty}\right\}.
\end{align}
The effectiveness of transferring the confounding effects from sources to the target by \eqref{eq:bias_correction_term} requires only mild assumptions about the confounding structure $f(\cdot)$. This also makes the profiled residual transfer fairly general and capable of addressing various confounding structures, rather than only the commonly assumed linear form.

\begin{remark}\label{emp:cate}
In many scenarios, the model shift $\eta_0$ carries its own scientific significance and may be our primary focus. For instance, in treatment effect estimation, the presence of unmeasured confounders frequently induces substantial bias in the resulting estimates. Existing methods often utilize instrumental variables or proxy methods to mitigate confounding effectss. Instead, if we regard the treatment group with indicator $T=1$ as the target model \eqref{eq:target_model}, and the control group with $T=0$ as the source model \eqref{eq:source_model}, the conditional average treatment effect $\tau(x)$ for a given covariate $x$ is indeed 
$\tau(x) = \mathbb{E}[Y|x,T = 1] - \mathbb{E}[Y|x,T = 0] = (\beta_t -\beta_s)^\top x$, which can be obtained once $\eta_0 = \beta_t - \beta_s$ is estimated.
\end{remark}

In applications, further investigation of the target parameter $\beta_t$ in \eqref{eq:target_model} is often needed. Given that $\hat{\eta}$ is already confounding-free, a natural estimator for $\beta_t$ is $\hat{\beta}_t = \hat{\beta}_s +\hat{\eta}$, where $\hat{\beta}_s$ is a deconfounded estimator for the source parameter $\beta_s$. For example, the estimator $\hat{\beta}_s$ can be obtained by applying trimmed regression to the pooled source data. We define the pooled  design matrix as $X_s = (X_s^{(1)\top}, \cdots, X_s^{(K)\top})^\top \in \mathbb{R}^{n_s \times p}$, the response vector as $Y_s = (Y_s^{(1)\top}, \ldots, Y_s^{(K)\top})^\top \in \mathbb{R}^{n_s}$ and define $\tilde{X}_s = \mathcal{Q}_s X_s$ and $\tilde{Y}_s = \mathcal{Q}_s Y_s$ as the covariates and response adjusted using the trim transform $\mathcal{Q}_s$ of the source data, as described similarly in \eqref{eq:trim_transform}. The estimator $\hat{\beta}_s$ is then given by
\begin{align}
\label{eq:estimator_betas}
    \hat{\beta}_s = \argmin_{\beta \in \mathbb{R}^p} \left\{\frac{1}{n_s}  \left\|\tilde{Y}_s - \tilde{X}_s \beta\right\|_2^2 +\lambda_s\|\beta\|_1 \right\}.
\end{align}
A greater concentration of confounding effects makes the trim transform more effective in obtaining the deconfounded estimator $\hat{\beta}_s$ in source models. 
\red{In addition to the trim method, the LAVA approach proposed by \cite{chernozhukov2017lava} and the synthetic two-stage regularized regression introduced by \cite{tang2023synthetic} can also be employed to obtain the source estimator for $\beta_s$.}
Mitigating confounding bias and hence obtaining reliable deconfounded estimators in source models is generally more feasible due to the large source sample size.

In summary, we address the challenge of hidden confounders in the target model via two steps. The first step involves estimating the model shift from source domains to the target domain using the proposed profiled residual transfer \eqref{eq:bias_correction_term}, which is robust to confounding structures and the initial estimates of the source models. The second step is the deconfounded estimation of the source models, which becomes more manageable due to the large sample size of the source data.
We refer to the entire approach as profiled transfer learning (ProTrans), and summarize the procedure based on the trim method in Algorithm \ref{algo:transfer_method}. \red{ This algorithm can also be directly extended to methods such as LAVA and the synthetic IV method by correspondingly modifying the source estimation procedure. }

  \begin{breakablealgorithm}\label{algo}
  \renewcommand{\algorithmicrequire}{\textbf{Input:}}
  \renewcommand{\algorithmicensure}{\textbf{Output:}}
  \caption{Profiled Transfer Learning Algorithm}\label{algo:transfer_method}
  \begin{algorithmic}[1]
  \REQUIRE Source data $(X^{(k)}_s,Y^{(k)}_s),k=1,\cdots,K$ with total sample size $n_s$ and target data $(X_t,Y_t)$ with sample size $n_t$. 
  \ENSURE Estimators for model shift $\eta_0$ and target parameter $\beta_t$. 
  
   \textbf{Source Parameter Estimation:}
  \STATE Singular value decomposition: $X_s= U_sD_sV_s$ where $X_s=(X_s^{(1)\top},\cdots,X_s^{(K)\top})^\top$;
  \STATE Trimming: $\mathcal{Q}_s \leftarrow U_s \operatorname{diag}\left(\min(\tau_s/D_s,1)\right)U^\top_s$, where $\tau_s$ can be taken as the $\lfloor n_s/2\rfloor$th ordered singular value of $X_s$;
  \STATE Trim transform: $\tilde{X}_s \leftarrow \mathcal{Q}_sX_s,\tilde{Y}_s \leftarrow \mathcal{Q}_sY_s$ where $Y_s = (Y_s^{(1)\top},\cdots,Y_s^{(K)\top})^\top$;
  \STATE LASSO estimation: Obtain $\hat{\beta}_s$ through \eqref{eq:estimator_betas};
 
  \textbf{Model Shift Estimation:}
  \STATE Initial estimation: Find some initial estimator $\hat{\beta}_{\text{init}}$ for $\beta_s$. 
  \STATE Profiled source residual: $\hat{Z}_s^{(k)} \leftarrow Y^{(k)}_s - X^{(k)}_s\hat{\beta}_{\text{init}},\quad 1\le k\le K$;
  \STATE Profiled residual transfer: Obtain the profiled target residual $\hat{Z}_t$ via \eqref{eq:bias_correction_term};
  \STATE LASSO estimation: Obtain $\hat{\eta}$ through \eqref{eq:estimator_of_eta};

  \textbf{Target Parameter Estimation:}
  \STATE Target estimation: $\hat{\beta}_t \leftarrow \hat{\beta}_s +\hat{\eta}$;
  \RETURN Estimators $\hat{\eta}$ of $\eta_0$ and $\hat{\beta}_t$ of $\beta_t$.
  \end{algorithmic}  
  \end{breakablealgorithm}

We conclude this section with the following remark which states that the direct application of transfer learning to confounded models fails to eliminate confounding-induced bias in model shift and target parameter estimation.
\begin{remark}
\label{remark:classical_transfer_learning}
 Based on the deconfounded source parameter estimator $\hat{\beta}_s$, the trimmed covariates $\tilde{X}_t$ and trimmed response $\tilde{Y}_t$, conventional transfer learning frameworks guide us to adopt the target estimator $\tilde{\beta}_t = \hat{\beta}_s +\tilde{\eta}$, where $\tilde{\eta}$ is an estimator for model shift such as the LASSO estimator for regressing $\tilde{Y}_t - \tilde{X}_t\hat{\beta}_s$ against $\tilde{X}_t$, i.e.,
\begin{align}
\label{eq:classical_estimator_eta}
    \tilde{\eta} = \argmin_{\eta\in \mathbb{R}^p } \left\{\frac{1}{n_t}\left\|\tilde{Y}_t - \tilde{X}_t\hat{\beta}_s - \tilde{X}_t\eta \right\|_2^2 +\tilde{\lambda}_t \|\eta\|_1\right\}. 
\end{align}
However, the bias term $b_t$ remains in the response of the target model:
\begin{align*}
    \tilde{Y}_t - \tilde{X}_t\hat{\beta}_s = \tilde{X}_t \left[\left(\beta_t - \beta_s\right) + \left(\beta_s - \hat{\beta}_s\right)+ b_t\right] + \tilde{U}_t + \tilde{\varepsilon}_t,
\end{align*}
so that $\tilde{\eta}$ still suffers from the bias introduced by hidden confounders, which persists regardless of the amount of available source data. Rigorous comparison of non-asymptotic error bounds between $\tilde{\eta}$ in \eqref{eq:classical_estimator_eta} and the ProTrans estimator $\hat{\eta}$ in \eqref{eq:estimator_of_eta} is presented in the next section. 
\end{remark}


\section{Theoretical Guarantees for ProTrans}
\label{sec:theorem}
\subsection{Model shift estimation}

We begin by analyzing the properties of the estimator $\hat{\eta}$ for $\eta_0 = \beta_t - \beta_s$. First, we define the parameter space for the source parameters $\beta_s^{(k)}$ and the target parameter $\beta_t$ as 
\begin{align*}
    \Theta(s_\delta,s_\eta) = \left\{\beta_s^{(k)},\beta_t\ \big|\ \|\beta_s^{(k)} - \beta_s\|_1 \le s_\delta, \|\eta_0\|_0\le s_\eta, k=1,2,\cdots,K\right\}. 
\end{align*}
In this section, we consider the case where $s_\delta$ is a constant and $s_\eta = o(n_t^{1/2})$. 
This parameter space requires that the difference between each source parameter $\beta_s^{(k)}$ and the average $\beta_s$ is bounded, and that the model shift $\eta_0$ is sparse. This condition is common in high-dimensional regression when transferring parameters from source to target. We denote the support set of $\eta_0$ as $S_\eta$ and its complement as $S_\eta^c$. The following standard assumption is imposed.


\begin{assump}
    \label{assum:distribution_E_H_f}
    \begin{enumerate}
         \item[a.] The errors $E_{ti},E_{si}^{(k)}$, the confounders $H_{ti},H_{si}^{(k)}$ and $f(H_{ti}),f(H_{si}^{(k)})$ are i.i.d. sub-Gaussian random variables. 
         \item[b.] The noise terms $\varepsilon_{ti},\varepsilon_{si}^{(k)}$ are i.i.d. sub-Gaussian random variables with zero mean and variance $\sigma^2$. 
        \item[c.] In the convex cone $\mathcal{A}_{s_\eta} = \left\{\eta\in \mathbb{R}^p; 3\|\eta_{S_\eta}\|_1 - \|\eta_{S_\eta^c}\|_1\ge 0\right\}$, the restricted eigenvalue condition holds for some constant $C_r>0$, i.e., $n_t^{-1}\left\|X_t\eta\right\|_2^2 \ge C_r \left\|\eta\right\|_2^2$ for $\eta\in \mathcal{A}_{s_\eta}$.
        \item[d.] The confounders exert the similar effects on the target and source models, i.e., 
        \begin{align*}
            \left\|\frac{1}{n_s}\sum_{k=1}^K n_s^{(k)}\phi_s^{(k)} - \phi_t\right\|_2 \le C \sqrt{\frac{\log p}{n_t}},
        \end{align*} 
        where $C$ is a constant. 
    \end{enumerate}
\end{assump}

Unlike the commonly used linear confounding structure $X = H\Psi + E$ for some $\Psi\in \mathbb{R}^{q\times p}$, our model assumes a more flexible structure: $X = f(H) + E$. In Assumption \ref{assum:distribution_E_H_f}.a, we only require that the random vectors $f(H)$ are sub-Gaussian. This condition encompasses not only the linear case $f(H) = H\Psi$ but also a broad class of functions. Assumptions \ref{assum:distribution_E_H_f}.b and \ref{assum:distribution_E_H_f}.c are standard assumptions in high-dimensional penalized regression. Assumption \ref{assum:distribution_E_H_f}.d requires that the average confounding effect in the source models is close to $\phi_t$ in the target model. This condition is essential for transferring confounding information to the target model appropriately. When each source is informative such that $\phi_s^{(k)}$ is close to $\phi_t$, the condition is clearly satisfied. In Section \ref{sec:source_selection}, we further relax this assumption to accommodate more general cases, where the confounding effect in at least one source model is similar to that in the target model, while others may differ significantly. \textcolor{black}{We remark that the theoretical guarantees of the ProTrans estimator for model shift require no specific assumptions about the true confounding biases in either the source or target domains, in contrast to most deconfounding methods.}

Under these standard assumptions, Theorem \ref{thm:upper_bound_of_eta} below provides non-asymptotic error bounds for the estimated model shift $\hat{\eta}$.
\begin{thm}
\label{thm:upper_bound_of_eta}
    Under Assumption \ref{assum:distribution_E_H_f}, suppose the initial source estimator $\hat{\beta}_{\text{init}}$ for $\beta_s$ satisfies $\|\hat{\beta}_{\text{init}} - \beta_s\|_1 \le C_s$ for a constant $C_s$. Then, 
    with probability at least $1-p^{-c}$ for a constant $c$, the estimator $\hat{\eta}$ defined in \eqref{eq:estimator_of_eta} satisfies:
    \begin{align*}
       \left\|\hat{\eta} - \eta_0\right\|_1 \le \frac{6}{C_r}s_\eta\lambda_t,\quad
        \left\|\hat{\eta} - \eta_0\right\|_2 &\le \frac{3}{2C_r}\sqrt{s_\eta}\lambda_t,\quad \frac{1}{\sqrt{n_t}}\left\|X_t\left(\hat{\eta} - \eta_0\right)\right\|_2 \le \frac{3}{2\sqrt{C_r}}\sqrt{s_\eta}\lambda_t.
    \end{align*}
    Here the tuning parameter $\lambda_t$ satisfies $
        \lambda_t\ge 4\left(C_1\sqrt{q}\|\phi_t\|_2 +C_2\right)\sqrt{\log p/n_t}$, where $q$ is the dimension of confounders and $C_1,C_2$ are constants.  
\end{thm}

In Theorem \ref{thm:upper_bound_of_eta}, by selecting the tuning parameter $\lambda_t\asymp \sqrt{\log p/n_t}$, the estimator $\hat{\eta}$ achieves the optimal convergence rate $\sqrt{\log p/n_t}$, even in the presence of hidden confounding. By introducing the profiled target residual $\hat{Z}_t$, ProTrans effectively removes the bias caused by confounders. It is also noteworthy that the result in Theorem \ref{thm:upper_bound_of_eta} is general and applies to many scenarios, given the mild conditions on the initial estimator $\hat{\beta}_{\text{init}}$ and the confounding structure $f(\cdot)$. 
As shown in Remark \ref{emp:cate}, Theorem \ref{thm:upper_bound_of_eta} can be directly used to estimate the conditional average treatment effect $\tau(x)$ in the presence of hidden confounders. When using ProTrans, the estimated conditional average treatment effect for a given covariate vector $x$ achieves root-$n$ convergence, provided that $\|x\|_{\infty}$ is bounded, even in the presence of hidden confounders.

Furthermore, we provide the minimax lower bound for estimating $\eta_0$. Let $\mathcal{B}_0(s_\eta)$ denote the $l_0$-norm ball $\mathcal{B}_0(s_\eta) = \left\{\eta\in\mathbb{R}^p; \|\eta\|_0 \le s_\eta\right\}$. 
\begin{prop}
    \label{prop:lower_bound_of_eta}
    Under the same conditions as in Theorem \ref{thm:upper_bound_of_eta}, it holds that
    \begin{align*}
       \inf_{\hat{\eta}} \sup_{\eta \in \mathcal{B}_0(s_\eta)} \mathbb{P}\left(\|\hat{\eta} - \eta_0\|_2 \ge c_0\sqrt{\frac{s_\eta \log p}{n_t}}\right) \ge \frac{1}{2},
    \end{align*}
    for some constant $c_0>0$. 
\end{prop}

Proposition \ref{prop:lower_bound_of_eta} shows that the proposed estimator $\hat{\eta}$ achieves the optimal convergence rate. For comparison, we also provide the estimation error of the estimator $\tilde{\eta}$ defined in Remark \ref{remark:classical_transfer_learning} which directly applies conventional transfer learning frameworks to confounded models. We show that $\tilde{\eta}$ is highly sensitive to hidden confounders, even under the simplest linear confounding structure $f(H)=H\Psi$. 
 Existing results \citep{cevid2020spectral,guo2022doubly} provide the following upper error bound for $\tilde{\eta}$:
\begin{align*}
    \|\tilde{\eta} - \eta_0\|_1 \lesssim \sqrt{\frac{p\|\phi_t\|_2^2}{n_t\lambda^2_q(\Psi)}}\vee\sqrt{\frac{\log p}{n_t}}  + \|\beta_s - \hat{\beta}_{s}\|_1, 
\end{align*}
 where $\lambda_q(\Psi)$ denotes the smallest singular value of $\Psi$. We also show the minimax lower bound for any possible estimator from the classical transfer learning methods. With slight abuse of notation, let $\tilde{\beta}_s$ be a generic estimator for $\beta_s$ using source data, $\tilde{\eta}$ be a generic estimator for the model shift $\beta_t-\beta_s$ using the transferred target data $(Y_t-X_t\tilde{\beta}_s,X_t)$. Further denote $\mathcal{F}_\eta$ to be the collection of such $\tilde{\eta}$. Proposition \ref{prop:lower_bound_of_baseline_eta_only} provides the minimax lower bounds for such $\tilde{\eta}\in \mathcal{F}_\eta$.
\begin{prop}
     \label{prop:lower_bound_of_baseline_eta_only}
     Under the same conditions as in Theorem \ref{thm:upper_bound_of_eta}, we have 
     \begin{align*}
       \inf_{\tilde{\eta}\in \mathcal{F}_\eta} \sup_{\eta \in \mathcal{B}_0(s_\eta)} \mathbb{P}\left(\|\tilde{\eta} - \eta_0\|_2 \ge c_1 \sqrt{\frac{p\|\phi_t\|_2^2}{n_t\lambda_q^2(\Psi)}} \vee c_2\sqrt{\frac{s_\eta\log p}{n_t}} \right)\ge \frac{1}{2},
     \end{align*}
     for some constants $c_1,c_2$.
\end{prop}
 
If the confounding effect $\|\phi_t\|_2$ is large, or the confounding structure is poorly conditioned (i.e., $\lambda_q(\Psi)$ is small), the term $\sqrt{p\|\phi_t\|_2^2/n_t\lambda^2_q(\Psi)}$ due to confounding bias might dominate the error bound, resulting in a slower convergence rate for $\tilde{\eta}$ compared to $\hat{\eta}$. Assuming $\|\phi_t\|_2$ is bounded, the confounding bias in $\tilde{\eta}$ remains small only if $\lambda_q(\Psi)\gg p^{1/2}\log^{-1/2}p$. This requirement is stringent, as it holds only when all nonzero singular values of $\Psi$ are of similar magnitude. Additionally, hidden confounders in the source models may make accurate estimation of $\beta_s$ infeasible. A large $\|\hat{\beta}_s - \beta_s\|_1$ can further increase the estimation error of $\tilde{\eta}$. In contrast, ProTrans does not rely on the linear confounding structure or a large $\lambda_q(\Psi)$, and only requires a bounded initial estimator $\hat{\beta}_{\text{init}}$ for $\beta_s$. The optimal convergence rate for $\hat{\eta}$ in Theorem \ref{thm:upper_bound_of_eta} remains valid even when the confounding structure is complex and the source and target models are heavily affected by hidden confounders. This represents a significant improvement of ProTrans over classical transfer learning estimators.

\begin{remark}
The homogeneity requirements in Assumptions \ref{assum:distribution_E_H_f}.a and \ref{assum:distribution_E_H_f}.b can be extended to the heterogeneous case, where the distributions of $H_{ti},E_{ti}$ are allowed to differ from those of $H_{si}^{(k)},E_{si}^{(k)}$. The heterogeneity affects the profiled residual transfer from $\hat{Z}_s$ to $\hat{Z}_t$. However, this issue can be effectively addressed if the initial estimator $\hat{\beta}_{\text{init}}$ converges rapidly. An intrinsic doubly robust property of the term $(X_s^\top X_s/n_s - X_t^\top X_t/n_t)(\hat{\beta}_{\text{init}} -\beta_s)$ in the profiled residual transfer-induced error ensures that the results in Theorem \ref{thm:upper_bound_of_eta} remain valid if either the covariance structures of the source and target models are sufficiently similar ($\|X_s^\top X_s/n_s - X_t^\top X_t/n_t\|_{\infty}$ is small) or the initial estimator $\hat{\beta}_{\text{init}}$ for $\beta_s$ is sufficiently accurate ($\|\hat{\beta}_{\text{init}} -\beta_s\|_1$ is small).  In the heterogeneous case, it suffices for the initial estimator to satisfy the rate $\left\|\hat{\beta}_{\text{init}}-\beta_s\right\|_1\lesssim \sqrt{\log p/n_t}$, which is likely to hold when $\hat{\beta}_{\text{init}}$ is chosen as the deconfounded estimator for the source models due to the relatively large sample size $n_s$ compared to $n_t$. We will present a detailed demonstration of the deconfounded estimator in the next section.
\end{remark}

\subsection{Target parameter estimation}\label{theory:target}
In this section, we consider the ProTrans estimator $\hat{\beta}_t = \hat{\beta}_s + \hat{\eta}$ for the target parameter $\beta_t$. Compared with the model shift estimator $\hat{\eta}$, the theoretical analysis of $\hat{\beta}_t$ additionally requires $\hat{\beta}_s$ to converge at a relatively fast rate. Suppose that the parameter $\beta_s$ has a support set $S_s$ with cardinality at most $s=o(n_s^{1/2})$. Although hidden confounders may bias the estimation of $\beta_s$, the large sample sizes in the source models generally provide sufficient accuracy in $\hat{\beta}_s$ when using the trim transform. To begin, we define the covariance matrix of $X_s$ as $\Sigma_X = \mathbb{E}[X_s^\top X_s/n_s]$ and impose the following regularity condition for the trim transform. 
 \red{
 \begin{assump}
     \label{assum:trim_transform}
     \begin{enumerate}
         \item[a.] The threshold $\tau_s$ in the trim transform $\mathcal{Q}_s$ defined in \eqref{eq:trim_transform} satisfies that $\tau_s \asymp p^{1/2}n_s^{-1/2}$. 
         \item[b.] The $\tau$-th eigenvalue and eigengap of $\Sigma_X = \mathbb{E}[X_s^\top X_s/n_s]$ is large enough, i.e., 
         \begin{align*}
             \lambda_{\tau}(\Sigma_X) \gg p^{1/2} \log^{3/2} p ,\quad \Delta_\tau = \lambda_\tau(\Sigma_X) - \lambda_{\tau+1}(\Sigma_X) \gg p^{1/2}n_s^{-1/4}\log^{3/4} p.
         \end{align*} 
         \item[c.] The restricted eigenvalue condition holds for some constant $C_r>0$ in the convex cone $\mathcal{A}_{S_s} = \left\{\beta \in \mathbb{R}^p; 3\|\beta_{S_s}\|_1 - \|\beta_{S_s^c}\|_1\ge 0\right\}$,i.e., $n_s^{-1}\|\mathcal{Q}_sX_s\beta\|_2^2 \ge C_r \|\beta\|_2^2$ for $\beta\in \mathcal{A}_{S_s}$. 
     \end{enumerate}
 \end{assump}
 }
The conditions in Assumption \ref{assum:trim_transform} regularize the trim transform. Assumption \ref{assum:trim_transform}.a imposes a standard requirement that the eigenvalues of the trimmed covariance matrix do not exceed $p^{1/2}n_s^{-1/2}$, which aligns with the typical bulk eigenvalues of a non-confounded empirical covariance matrix.
Assumption \ref{assum:trim_transform}.b requires that both the $\tau$-th eigenvalue and the eigengap of $\Sigma_X$ are sufficiently large. The condition commonly used to eliminate hidden confounders requires that the top eigenvalues of $\Sigma_X$ are on the order of $p$. A weaker condition for consistent estimation of $\beta_s$ is that the top eigenvalues of $\Sigma_X$ are on the order of $pn_s^{-1}$. Compared to these, our condition on the top eigenvalues of $\Sigma_X$ is mild because we only require the confounding structure $f(\cdot)$ to induce spikes of order $p^{1/2}\log^{3/2} p$ in the covariance matrix $\Sigma_X$. The eigengap condition is also mild because the bulk eigenvalues of $\Sigma_X$ are constant. The presence of spikes further ensures this condition, especially in source models where $n_s$ is larger than in the target model.

Theorem \ref{thm:upper_bound_of_betas} provides the error bounds for the estimator $\hat{\beta}_s$ from multiple source datasets. 
 \begin{thm}
     \label{thm:upper_bound_of_betas}
     Under Assumptions \ref{assum:distribution_E_H_f} and \ref{assum:trim_transform}, with probability at least $1-p^{-c}$, the source estimator $\hat{\beta}_s$ defined in \eqref{eq:estimator_betas} satisfies 
     \begin{align}
     \label{eq:thm:source_bound}
            \|\hat{\beta}_s - \beta_s\|_1  \le C_3 s\lambda_s + \frac{1}{\lambda_s} \frac{p\|\phi_t\|_2^2}{n_s \lambda_\Psi^2},\quad \|\hat{\beta}_s - \beta_s\|_2  \le C_3 \sqrt{s}\lambda_s + \frac{1}{\lambda_s}\frac{p\|\phi_t\|_2^2}{n_s \lambda_\Psi^2},
        \end{align}
        where the tuning parameter $\lambda_s$ satisfies $\lambda_s \ge C_4 \sqrt{\log p/n_s} +C_5 \sqrt{p/n_s}\lambda_\Psi^{-1}\|\phi_t\|_2$, $\lambda^{-1}_\Psi$ denotes the largest singular value of $\Sigma_X^{-1}\mathbb{E}[f^\top(H_s)H_s/n_s]$ and $c,C_3,C_4,C_5$ are constants. 
 \end{thm}



Theorem \ref{thm:upper_bound_of_betas} addresses general multi-source models with unknown nonlinear confounding structures. In the special case of a single-source model with a linear confounding structure $X_s = H_s\Psi +E_s$, the results in Theorem \ref{thm:upper_bound_of_betas} align with Theorem 2 in \cite{cevid2020spectral}. The error bound \eqref{eq:thm:source_bound} for $\hat{\beta}_s$ consists of two components: the first term represents the standard Lasso estimation error when all confounding effects have been accounted for, and the second term is the confounder-induced error, given by $p\|\phi_t\|_2^2/(\lambda_s n_s \lambda_\Psi^2)$. The parameter $\lambda_\Psi$ indicates how concentrated the confounding effects are and a large $\lambda_\Psi$ reduces this bias term. When the confounding effects $H_s^{(k)}\phi_s^{(k)}$ are primarily concentrated in the leading eigenspaces of $\Sigma_X$, the trim transform $\mathcal{Q}_s$ with appropriate choices of $\tau$ and $\tau_s$ satisfying Assumption \ref{assum:trim_transform}, shrinks confounding effects along directions associated with large singular values, thereby significantly reducing the confounding bias. A large $\lambda_s$ can also reduce the bias term, as the stronger $l_1$-penalty helps to shrink the bias $b_s$, although it is not required to be $l_1$-sparse. 


It is notable that $\lambda_\Psi$ reflects the population level properties of the confounding structure. Specifically, $\lambda_\Psi$ measures the spike eigenvalues of $\Sigma_X$ caused by the confounding structure $f(\cdot)$ which remains consistent in both the source and target models. However, applying the trim transform and the deconfounded estimator to the source models is still more practical and effective than directly deconfounding the target model. In high-dimensional settings, the empirical covariance matrix can deviate significantly from $\Sigma_X$ and the sample size greatly influences whether these spikes can be accurately detected. Due to the large sample size $n_s$ of source models, these spikes are more likely to appear prominently in the empirical covariance matrices $X_s^\top X_s/n_s$ of source models than in the target model. The concentration of confounding effects in the leading eigenspaces of $X_s^\top X_s/n_s$ is governed by the eigengap ratio between the spike and bulk eigenvalues of $X_s^\top X_s/n_s$, given by $\lambda_\Psi^2/(p/n_s)$. This term also appears in the confounder-induced error in Theorem  \ref{thm:upper_bound_of_betas}. In contrast, the smaller sample size $n_t$ in the target model makes it more difficult for the empirical covariance matrix $X_t^\top X_t/n_t$ to capture the spikes in $\Sigma_X$. 

Combining the source estimator $\hat{\beta}_s$ for $\beta_s$ with the model shift estimator $\hat{\eta}$ for $\eta_0$, we obtain the estimation error for the target estimator $\hat{\beta}_t$ in Theorem \ref{thm:upper_bound_of_betat}. 

\begin{thm}
    \label{thm:upper_bound_of_betat}
    Under the same conditions as in Theorems \ref{thm:upper_bound_of_eta} and \ref{thm:upper_bound_of_betas}, with probability at least $1-p^{-c}$, the proposed estimator $\hat{\beta}_t$ for $\beta_t$ satisfies
         \begin{align*}
            \|\hat{\beta}_t - \beta_t\|_1  &\le  C_6 \sqrt{\frac{p\|\phi_t\|_2^2}{n_s\lambda_\Psi^2}} \vee C_7 s_\eta\sqrt{\frac{\log p}{n_t}},\\
            \|\hat{\beta}_t - \beta_t\|_2  &\le  C_6 \sqrt{\frac{p\|\phi_t\|_2^2}{n_s\lambda_\Psi^2}} \vee C_7\sqrt{\frac{ s_\eta\log p}{n_t}},
        \end{align*}
        where we set $\lambda_t=\sqrt{\log p/n_t},\lambda_s = \sqrt{\log p/n_s}\vee \lambda_\Psi^{-1}\|\phi_t\|_2\sqrt{p/n_s}$ and $c,C_6,C_7$ are constants. 
\end{thm}

In Theorem \ref{thm:upper_bound_of_betat}, the estimation error of the ProTrans estimator $\hat{\beta}_t$ consists of two components: the confounding-free error in estimating $\eta_0$ and the source estimation error in $\beta_s$. The confounding effect appears only in the source estimation and can be mitigated due to the large sample size of the source data. In the following proposition, we show that the ProTrans estimator $\hat{\beta}_t$ also attains the minimax lower bound for this problem. 
\begin{prop}
    \label{prop:minimax_for_beta_t}
    Under the same conditions as in Theorems \ref{thm:upper_bound_of_eta} and \ref{thm:upper_bound_of_betas}, we have that for any estimator $\hat{\beta}_t$ of $\beta_t$,
     \begin{align*}
     \underset{\substack{\beta_t = \beta_s + \eta \\ \eta \in \mathcal{B}_0(s_\eta),\beta_s\in \mathcal{B}_0(s)}}{\sup} \mathbb{P}\left(\|\hat{\beta}_t - \beta_t\|_2 \ge c_1 \sqrt{\frac{p\|\phi_t\|_2^2}{n_s\lambda_\Psi^2} }\vee c_2\sqrt{\frac{s_\eta\log p}{n_t}} \right)\ge \frac{1}{2},
     \end{align*} 
     for some constants $c_1,c_2$. 
\end{prop}

Meanwhile, we also establish the minimax result for estimators of $\beta_t$ when directly applying the conventional transfer learning method to confounded models. Recall the definition of $\tilde{\beta}_s,\tilde{\eta}$ in Proposition \ref{prop:lower_bound_of_baseline_eta_only} and define the set $\mathcal{F}_\beta$ as the collection of all classical transfer learning estimators of $\beta_t$ in the form $ \tilde{\beta}_t = \tilde{\beta}_s +\tilde{\eta}$.
\begin{prop}
     \label{prop:lower_bound_of_baseline_eta}
     Under the same conditions as in Theorems \ref{thm:upper_bound_of_eta} and \ref{thm:upper_bound_of_betas}, we have 
     \begin{align*}
       \inf_{\tilde{\beta}_t\in \mathcal{F}_\beta}\underset{\substack{\beta_t = \beta_s + \eta \\ \eta \in \mathcal{B}_0(s_\eta),\beta_s\in \mathcal{B}_0(s)}}{\sup}\mathbb{P}\left(\|\tilde{\beta}_t - \beta_t\|_2 \ge c_1 \sqrt{\frac{p\|\phi_t\|_2^2}{n_t\lambda_\Psi^2} }\vee c_2\sqrt{\frac{s_\eta\log p}{n_t}} \right)\ge \frac{1}{2},
     \end{align*}
     for some constants $c_1,c_2$. 
\end{prop}



When the confounding structure $f(\cdot)$ is difficult to identify ($\lambda_\Psi$ is small) or the confounding effect is large ($\|\phi_t\|_2$ is large), the confounder-induced bias dominates the lower bound of $\tilde{\beta}_t$. Classical transfer learning methods cannot reduce this bias, regardless of the size of the source sample. As shown in Proposition \ref{prop:lower_bound_of_baseline_eta}, the confounding bias remains at $n_t^{-1/2}p^{1/2}\|\phi_t\|_2\lambda^{-1}_\Psi$. In contrast, ProTrans significantly reduces the confounding bias to $n_s^{-1/2}p^{1/2}\|\phi_t\|_2\lambda^{-1}_\Psi$ as shown in Theorem \ref{thm:upper_bound_of_betat}. With ProTrans, the larger source sample size helps mitigate the error caused by hidden confounding. Consequently, the condition for achieving a sufficiently small confounding bias in $\hat{\beta}_t$ is relaxed to $\lambda_\Psi \gg (n_t/n_s)^{1/2} p^{1/2}\log^{-1/2}p$. This requirement is much milder when the source sample size is significantly larger than the target sample size.

\red{\begin{remark}
The theoretical results and associated benefits are not tied to the trim method used for source estimation in the high-dimensional setting. The same profiled residual transfer idea can also be combined with other deconfounding methods, such as LAVA \citep{chernozhukov2017lava} and synthetic two-stage regularized regression \citep{tang2023synthetic}, under both high- and fixed-dimensional scenarios. Analogous results are provided in the supplementary material.
\end{remark}}

\section{Informative Source Selection for ProTrans}
\label{sec:source_selection}
In Assumption \ref{assum:distribution_E_H_f}, we assume that the sources are informative such that the average of the confounding-effect coefficients $\phi_s^{(k)},1\le k\le K$ is close to $\phi_t$ in the target model. However, in many realistic scenarios, some sources may have confounding effects that deviate significantly from $\phi_t$, making them unlikely to help eliminate the confounding effect in the target model. 

We propose a source selection procedure to address this issue. First, we consider an initial estimator $\hat{\beta}^{(k)}_{\text{init}}$ of $\beta_s^{(k)}$. Let $\hat{Z}_t^{(k)}$ be the profiled residual term defined as in \eqref{eq:bias_correction_term}, and $\tilde{Z}_t^{(k)}=\mathcal{Q}_t\hat{Z}_t^{(k)}$. Further take the trimmed estimator of $\eta^{(k)}_0 = \beta_t - \beta_s^{(k)}$ as
\begin{align*}
\hat{\eta}^{(k)} = \argmin_{\eta\in \mathbb{R}^p}\left\{\frac{1}{n_t}\left\|\tilde{Y}_t - \tilde{X}_t \hat{\beta}^{(k)}_{\text{init}} - \tilde{Z}_t^{(k)} - \tilde{X}_t\eta \right\|_2^2 + \lambda_t^{(k)} \|\eta\|_1\right\}. 
\end{align*}
Here we use trimmed regression to ensure the stability and consistency of $\hat{\eta}^{(k)}$ even when $\phi_s^{(k)}$ differs much from $\phi_t$.
Next, we define the quantity
\begin{align*}
    v^{(k)} = \left\|\frac{X_t^\top}{n_t}\left(Y_t - X_t\hat{\beta}^{(k)}_{\text{init}} - \hat{Z}^{(k)}_{t} - X_t\hat{\eta}^{(k)}\right)\right\|_{\infty}.
\end{align*}
Intuitively, when $\phi_s^{(k)}$ is close to $\phi_t$, the profiled residual $\tilde{Z}_t^{(k)}$ can transfer the confounding information from the $k$-th source to the target model. Consequently, the estimator $\hat{\eta}^{(k)}$ remains accurate and $v^{(k)}$ is small. In contrast, when $\phi_s^{(k)}$ differs substantially from $\phi_t$, the transfer process introduces a term proportional to the discrepancy between the confounding effects of the $k$-th source model and the target model. This discrepancy enlarges $v^{(k)}$ when the gap $\phi_s^{(k)}-\phi_t$ increases.

Based on this observation, we define 
$$\mathcal{I}(\rho) = \left\{k\ |\ v^{(k)} \le \rho\min_{1\le k\le K}v^{(k)}, \ k=1,2,\cdots, K \right\}$$
as the informative source group for some constant $\rho \ge 1$. Using only source data in $\mathcal{I}(\rho)$, we obtain the estimator $\hat{\beta}_t^{(\mathcal{I}(\rho))}$ via ProTrans. Here we consider the case that all source parameters $\beta_s^{(k)},1\le k\le K$ are transferable. Formally, the coefficient $\beta_s^{(k)}$ has a support set $S_s^{(k)}$ with $|S_s^{(k)}|\le s$ and the difference $\eta_0^{(k)} = \beta_t - \beta_s^{(k)}$ has a support set $S_\eta^{(k)}$ with $|S_\eta^{(k)}| \le s_\eta$ for $1\le k\le K$. This helps us focus on detecting sources containing informative confounding structures. If some source models are non-transferable due to large or dense coefficients $\eta_0^{(k)}$, existing source selection methods can be applied beforehand \citep{li2022transfer}. To study the theoretical properties of $\hat{\beta}_t^{(\mathcal{I}(\rho))}$, we introduce the following assumption.
\begin{assump}
\label{assum:source_detection}
    \begin{enumerate}
        \item[a.] For each $k=1,\ldots,K$, the sample size $n_s^{(k)}$ is larger than $n_t$. 
        \item[b.] It holds that $\lambda_\Psi \ge C_{\lambda} p^{1/2}n_t^{-1/2}$, and there exists at least one informative source model in the sense that there exists some $k, \ 1\le k\le K$, such that $\|\phi_s^{(k)}-\phi_t\|^2_2 \le C_{\phi}\lambda_\Psi^2p^{-1}\log p$ for constants $C_{\lambda},C_{\phi}$. 
    \end{enumerate}
\end{assump}
Assumption \ref{assum:source_detection}.a requires each source dataset to have a larger sample size than $n_t$. This condition can be relaxed by aggregating smaller source datasets into a single one with a total sample size exceeding $n_t$. 
Assumption \ref{assum:source_detection}.b posits source-target model similarity, which guarantees the efficacy of ProTrans. This requirement is mild, as we only need at least one informative source model and make no demands on the characteristics of the other sources.
 Theorem \ref{thm:source_selection} provides the estimation error for $\hat{\beta}_t^{(\mathcal{I}(\rho))}$.
\begin{thm}
\label{thm:source_selection}
    Under Assumptions \ref{assum:distribution_E_H_f}, \ref{assum:trim_transform} and \ref{assum:source_detection}, with probability at least $1-p^{-c}$, the following inequalities hold:
    \begin{align*}
        \|\hat{\beta}_t^{(\mathcal{I}(\rho))}  - \beta_t\|_1 &\le C_8\sqrt{\frac{p\|\phi_t\|_2^2 }{n_s^{(\mathcal{I}(\rho))}\lambda_\Psi^2}}\vee C_9 (s_\eta +\rho)\sqrt{\frac{\log p}{n_t}},\\
        \|\hat{\beta}_t^{(\mathcal{I}(\rho))}  - \beta_t\|_2 &\le C_8\sqrt{\frac{p\|\phi_t\|_2^2 }{n_s^{(\mathcal{I}(\rho))}\lambda_\Psi^2}}\vee C_9 (\sqrt{s_\eta}+\rho)\sqrt{\frac{\log p}{n_t}},
    \end{align*}
    where $n_s^{(\mathcal{I}(\rho))} = \sum_{k\in \mathcal{I}(\rho)} n_s^{(k)}$ is the total sample size of the source data in the informative source group $\mathcal{I}(\rho)$, and $c,C_8,C_9$ are constants.              
\end{thm}

If $\|\phi_s^{(k)}-\phi_t\|_2^2 \le C_\phi  \lambda_\Psi^2 p^{-1} \log p$ for the $k$-th source, then $v^{(k)}$ falls, with high probability, within the interval $[c\sqrt{\log p/n_t},C\sqrt{\log p/n_t}]$ for some constants $c,C>0$. With a properly chosen constant $\rho$, the proposed selection procedure includes all informative source data in $\mathcal{I}(\rho)$. When all sources are informative, the results in Theorem \ref{thm:source_selection} are similar to those in Theorem \ref{thm:upper_bound_of_betat}, indicating that the source selection procedure does not result in a significant loss of efficiency. For a noninformative source model $k'$ where $\|\phi_s^{(k')} -\phi_t\|_2$ is large, the corresponding quantity $v^{(k')}$ will also be large; specifically, $v^{(k')}\gg \sqrt{\log p/n_t}$. Thus it will be excluded from the informative source group $\mathcal{I}(\rho)$. 
A small $\rho$ reduces the sample size $n_s^{(\mathcal{I}(\rho))}$ of source data used in transfer learning, increasing confounding bias in source estimation; a large $\rho$ results in greater discrepancies between $\phi_s^{(k)}$ and $\phi_t$ for $k \in \mathcal{I}(\rho)$, weakening the effectiveness of the profiled residual transfer. The optimal choice of $\rho$ is a common challenge in rank-based selection methods and is frequently discussed \citep{baranowski2020ranking,malhotra2021threshold}. Numerical results presented later indicate that the performance is not highly sensitive to the choice of $\rho$, allowing practitioners to select $\rho$ approximately based on prior knowledge or expert guidance.

\section{Simulation Study}
\label{sec:simulation}

In this section, we evaluate the performance of ProTrans under high-dimensional confounded models. We consider both linear and nonlinear confounding structures. Specifically, for the source and target models (denoted by the subscript $l$ as $s$ and $t$, respectively), the outcome model is given by
$Y_l = \beta_l^\top X_l + \phi_l^\top H_l + \varepsilon_l$, 
with three types of confounding structures as
\red{\begin{equation*}
   \begin{array}{ll}
\text{Scenario 1 (linear):}&  X_l = H_l^\top \Psi + E_l,\\
\text{Scenario 2 (nonlinear 1):}& X_l = 4\left(1+\exp(-H_l^\top \Psi)\right)^{-1} + 4\sin (H_l^\top \Psi) + E_l,\\
\text{Scenario 3 (nonlinear 2):} & X_l = 4\tanh\!\left(H_l^\top \Psi\right) + E_l.
\end{array} 
\end{equation*}}
The source and target data $(X_{li},Y_{li}), \ l=s,t, 1\le i \le n_l$ are generated from the above models. The hidden confounders $H_{li}$ are independently sampled from $\mathcal{N}(0, I_q)$, the random noise $\varepsilon_{li}$ from $\mathcal{N}(0,\sigma^2)$ and  $E_{li}$ from $\mathcal{N}(0,\sigma_E^2 I_p)$. The matrix $\Psi \in \mathbb{R}^{q\times p}$ is a random matrix with i.i.d. entries drawn from $0.5 n_s^{-1/2}\log n_s \mathcal{N}(0,1)$. The scaling factor $n_s^{-1/2}\log n_s$ ensures that the influence of the confounders $H_s$ is sufficiently strong for the source data. We set $q = 3$, $\sigma = 1$, and $\sigma_E = 2$ and we consider both the single-source and multi-source cases. In the single-source case, the source sample size is set to $n_s = 600$, and the target sample size to $n_t = 100$. The true source coefficient is defined as $\beta_s = 2\sum_{i=1}^5 e_i$, where $e_i\in \mathbb{R}^p$ is the vector whose $i$-th entry is equal to one and all other entries equal to zero. The target coefficient is given by $\beta_t = \beta_s + \eta_0$, where $\eta_0 = -4e_1 + 2 e_{p}$. In the multi-source case, the number of sources is set to $K = 4$, with a sample size $n_s^{(k)} = 150$ for $1 \le k \le 4$. The coefficient for the $k$-th source is defined as $\beta_s^{(k)} = 2\sum_{i=1}^5 e_i +e_k$ and hence $\beta_s = \sum_{k=1}^K \beta_s^{(k)}/4$. The values of $n_t$ and $\eta$ are identical to those in the single-source case.

We compare the performance of ProTrans with the following four baseline methods under the high-dimensional setting. 
\begin{itemize}
\item [(1)] \textit{SingleTrim}: applying the trimmed lasso method \citep{cevid2020spectral} only to the target data.
\item [(2)] \textit{TransTrim}: combining standard transfer learning \citep{li2022transfer} with a trimming step, resulting in $\tilde{\eta}$ in \eqref{eq:classical_estimator_eta}, the trimmed source estimator $\hat{\beta}_s$, and $\tilde{\beta}_t=\hat{\beta}_s+\tilde{\eta}$.
\item [(3)] \textit{SingleFarm}: adopting FARM \citep{fan2021robust} for deconfounding only the target data, in which the trim transform in the \textit{SingleTrim} is replaced by a projection onto the orthogonal complement of the factor space estimated using both the source and target data.
\item [(4)] \textit{TransFarm}: combining standard transfer learning with FARM.
\end{itemize}
\red{In addition, we evaluate the performance of ProTrans in the fixed-dimensional setting and the additional numerical results are provided in the supplementary material.} 

We set $\tau = \lfloor n_t/2\rfloor$ for the trim transform \eqref{eq:trim_transform} in the proposed method ProTrans and baselines {\it SingleTrim}, {\it TransTrim}. For {\it SingleFarm} and {\it TransFarm}, we define the factor space to be that spanned by top $r$ right singular vectors of $X = (X_t^\top,X_s^\top)^\top \in \mathbb{R}^{(n_s+n_t)\times p}$, where $r$ is chosen based on the largest relative eigen-gap, i.e., $r = \argmax_{1\le j\le p-1} (\sigma_j(X'X) - \sigma_{j+1}(X'X))/\sigma_{j+1}(X'X)$, with $\sigma_i(M)$ denoting the $i$-th eigenvalue of $M$. The tuning parameters $\lambda_s$ and $\lambda_t$ are selected through cross-validation using the standard function provided in the \textit{glmnet} package in R.

First, we evaluate the performance of ProTrans and the four baseline approaches across different covariate dimensions $p$. We set $\phi_s = \phi_t$, which are sampled from $\mathcal{N}(0,\sigma_\phi^2 I_p)$ with $\sigma_\phi = 1$ for Scenario 1 and $\sigma_\phi = 8$ for Scenario 2 and 3. Each simulation is repeated 100 times. The average $l_2$-norm errors in estimating $\eta$ and $\beta_t$ across all simulation runs are presented in Table \ref{table: simulation_p}. Across different values of $p$, we observe that ProTrans consistently outperforms the existing methods. In Scenario 1, due to the linear structure of the covariates $X = H^\top \Psi + E$, both the trim transform and orthogonal projection help to mitigate the confounding effects by estimating a subspace where hidden confounders exert minimal influence. As a result, while baseline methods still perform reasonably well,  our method does not rely on estimating the hidden confounder space and achieves higher accuracy. In Scenarios 2 and 3, the covariates follow a nonlinear and unknown structure, making it difficult to estimate the confounder space. This leads to substantial estimation biases for the baseline methods. In contrast, ProTrans remains effective.

\begin{table}[!ht]
\renewcommand{\arraystretch}{0.65}
\footnotesize
    \centerfloat
    \caption{The $l_2$ norm errors of different methods across various dimensions $p$.}
    \label{table: simulation_p}
    \begin{tabular}{cccccc}
        \toprule
        \multirow{2}{*}{$\bm{p}$} & \multirow{2}{*}{\textbf{Method}} & \multicolumn{2}{c}{\textbf{single-source}} & \multicolumn{2}{c}{\textbf{multi-source}}\\
        \cmidrule(lr){3-4} \cmidrule(lr){5-6} 
        & & $\eta$ & $\beta_{t}$ & $\eta$& $\beta_{t}$ \\
        \midrule
        \multicolumn{6}{c}{\textbf{Scenario 1 (linear)}}\\
        \hline
        \multirow{5}{*}{1000} 
        & SingleTrim  & $0.6095\ (0.0393)$ & $0.7685\ (0.0477)$ & $0.4756\ (0.0287)$ & $0.6057\ (0.0350)$ \\
        & TransTrim   & $0.4246\ (0.0247)$ & $0.3843\ (0.0220)$ & $0.3538\ (0.0174)$ & $0.3212\ (0.0148)$ \\
        & SingleFarm  & $1.0009\ (0.0490)$ & $1.4586\ (0.0609)$ & $1.0485\ (0.0723)$ & $1.5430\ (0.0865)$ \\
        & TransFarm   & $0.2824\ (0.0226)$ & $0.3365\ (0.0206)$ & $0.2493\ (0.0230)$ & $0.3112\ (0.0203)$ \\
        & ProTrans    & $\bm{0.2480}\ (0.0148)$ & $\bm{0.2422}\ (0.0129)$ & $\bm{0.2143}\ (0.0109)$ & $\bm{0.2160}\ (0.0088)$ \\
        \hline
        \multirow{5}{*}{2000} 
        & SingleTrim  & $0.8748\ (0.0657)$ & $1.0657\ (0.0772)$ & $0.6900\ (0.0522)$ & $0.8566\ (0.0612)$ \\
        & TransTrim   & $0.4725\ (0.0336)$ & $0.4270\ (0.0297)$ & $0.4105\ (0.0259)$ & $0.3738\ (0.0233)$ \\
        & SingleFarm  & $0.8936\ (0.0472)$ & $1.1931\ (0.0573)$ & $0.8374\ (0.0520)$ & $1.1614\ (0.0624)$ \\
        & TransFarm   & $0.3306\ (0.0209)$ & $0.3329\ (0.0168)$ & $0.3130\ (0.0194)$ & $0.3255\ (0.0163)$ \\
        & ProTrans    & $\bm{0.2894}\ (0.0173)$ & $\bm{0.2706}\ (0.0150)$ & $\bm{0.2213}\ (0.0106)$ & $\bm{0.2185}\ (0.0092)$ \\
        \hline
         \multicolumn{6}{c}{\textbf{Scenario 2 (nonlinear)}}\\
         \hline
        \multirow{5}{*}{1000} 
        & SingleTrim  & $4.0412\ (0.3030)$ & $4.8515\ (0.3597)$ & $2.1601\ (0.1799)$ & $2.7241\ (0.1683)$ \\
        & TransTrim  & $2.3566\ (0.1911)$ & $2.1424\ (0.1774)$ & $1.5994\ (0.1799)$ & $1.4360\ (0.1683)$\\
        & SingleFarm  & $8.1897\ (0.4476)$ & $9.8817\ (0.4883)$ & $10.806\ (0.5213)$ & $12.945\ (0.5661)$\\
        & TransFarm  & $2.5817\ (0.1799)$ & $2.5789\ (0.1683)$ & $3.5975\ (0.1853)$ & $3.5881\ (0.1732)$\\
        & ProTrans  & $\bm{0.6330}\ (0.0438)$ & $\bm{0.6231}\ (0.0400)$ & $\bm{0.5221}\ (0.0449)$ & $\bm{0.5329}\ (0.0455)$\\
        \hline
        \multirow{5}{*}{2000} 
        & SingleTrim  & $6.0142\ (0.5298)$ & $6.9928\ (0.6134)$ & $3.3560\ (0.1853)$ & $4.0193\ (0.1732)$ \\
        & TransTrim  & $3.0044\ (0.2569)$ & $2.7457\ (0.2364)$ & $2.0070\ (0.1853)$ & $1.7980\ (0.1732)$\\
        & SingleFarm   & $8.8878\ (0.5213)$ & $10.327\ (0.5661)$ & $11.576\ (0.1853)$ & $13.430\ (0.1732)$\\
        & TransFarm   & $2.8150\ (0.1853)$ & $2.7142\ (0.1732)$ & $4.3404\ (0.1853)$ & $4.1830\ (0.1732)$\\
        & ProTrans   & $\bm{0.6586}\ (0.0449)$ & $\bm{0.6497}\ (0.0455)$ & $\bm{0.5187}\ (0.1853)$ & $\bm{0.5206}\ (0.1732)$\\
        \hline
         \multicolumn{6}{c}{\textbf{Scenario 3 (nonlinear)}}\\
         \hline
        \multirow{5}{*}{1000} 
        & SingleTrim  & $6.5866\ (0.4950)$ &  $8.0162\ (0.6038)$  & $3.7007\ (0.2904)$& $4.6862\ (0.3561)$ \\
        & TransTrim  & $3.8329\ (0.3128)$& $3.4703\ (0.2874)$  & $2.6072\ (0.2139)$ & $2.3395\ (0.1888)$ \\
        & SingleFarm   & $3.7907\ (0.2981)$  & $5.1480\ (0.3473)$ & $5.3765\ (0.4486)$& $7.0284\ (0.5122)$ \\
        & TransFarm   & $1.4845\ (0.1335)$  & $1.5292\ (0.1169)$  & $1.8390\ (0.1596)$ & $1.8965\ (0.1483)$ \\
        & ProTrans    & $\bm{0.8309}\ (0.0761)$  & $\bm{0.8661}\ (0.0705)$  & $\bm{0.6457}\ (0.0518)$ & $\bm{0.6869}\ (0.0463)$ \\
        \hline
        \multirow{5}{*}{2000} 
        & SingleTrim  & $8.5861\ (0.6055)$  & $10.2241\ (0.7212)$  & $5.5476\ (0.4822)$ & $6.7159\ (0.5701)$ \\
        & TransTrim   & $4.8125\ (0.3914)$  & $4.3644\ (0.3549)$ & $3.2512\ (0.2560)$ & $2.9090\ (0.2298)$ \\
        & SingleFarm  & $4.1824\ (0.3275)$  & $5.2890\ (0.3820)$ & $5.7977\ (0.4710)$  & $7.1479\ (0.5304)$ \\
        & TransFarm   & $1.6215\ (0.1226)$ & $1.5747\ (0.1034)$ & $2.1802\ (0.1616)$  & $2.1088\ (0.1445)$ \\
        & ProTrans   & $\bm{0.8115}\ (0.0597)$ & $\bm{0.8480}\ (0.0625)$ & $\bm{0.6699}\ (0.0503)$ & $\bm{0.6975}\ (0.0503)$ \\
        \bottomrule
    \end{tabular}
\end{table}

Second, we examine the impact of varying confounding levels on ProTrans and the baseline methods by adjusting the norm of the confounder effects $\|\phi_t\|_2$ while keeping the dimensionality at $p = 1500$ and all other settings unchanged. As $\|\phi_t\|_2$ increases, the confounding effect becomes more severe, making it more challenging to accurately estimate both $\eta$ and $\beta_t$. ProTrans consistently outperforms the baselines and demonstrates greater robustness to increasing confounder levels in all settings. \red{Results for the $l_1$ and $l_2$ norm errors under various confounding levels $\|\phi_t\|_2$ and additional values of the dimension $p$ are reported in the supplementary material to save space.}

Lastly, we study the finite-sample performance of the proposed informative source selection procedure, where the confounding effect $\phi_s$ in the source models may differ from $\phi_t$ in the target model. In this simulation, we adopt the same multi-source setting as in the first experiment above, but choose $\phi_s^{(k)} = (5-k)\phi_t/4  + (k-1)\phi_t^{(k)}/4 $ for the $k$-th source model, $1\le k \le 4$. Here, $\phi_t^{(k)}$ denotes an i.i.d. copy of $\phi_t$. Table \ref{table: source_selection} presents the results of ProTrans after applying the source selection procedure with different levels of $\rho$, where $\rho =\infty$ corresponds to ProTrans without source selection. The proposed procedure identifies more informative sources and enhances transfer performance. The numerical results show a trade-off when choosing $\rho$, which is consistent with the findings in Theorem \ref{thm:source_selection}. 

\begin{table}[!ht]
\renewcommand{\arraystretch}{0.65}
\footnotesize
    \centerfloat
    \caption{The $l_2$ norm errors of source-selected methods under different selection level $\rho$.}
     \label{table: source_selection}
    \begin{tabular}{cccccccccc}
    \toprule
    \textbf{Scenario} & $\bm{p}$ & $\rho = 1$ & $\rho = 1.2$& $\rho = 1.4$ & $\rho = 1.6$ & $\rho = 2$ & $\rho = 2.5$ & $\rho = 3$ & $\rho=\infty$ \\
    \midrule
       \multirow{4}{*}{\textbf{Scenario 1}}  & 500 &  $0.3984$& $0.3785$ &$0.3541$&$0.3398$& $0.3356$ &$0.3400$ &$0.3392$& $0.3395$  \\
                                        & 1000 &  $0.4040$ &$0.3772$ &$0.3614$ &$0.3484$ &$0.3424$ &$0.3453$& $0.3479$& $0.3555$   \\ 
                                         & 1500 &  $0.4136$& $0.3861$ &$0.3654$ &$0.3524$ &$0.3476$ &$0.3536$ &$0.3545$& $0.3622$   \\ 
                                          & 2000 &  $0.4131$ &$0.3815$ &$0.3641$ &$0.3561$ &$0.3518$& $0.3552$& $0.3575$ &$0.3647$    \\ 
                                          \hline
       \multirow{4}{*}{\textbf{Scenario 2}}  & 500 &  $1.1866$ &$1.1467$& $1.1566$ &$1.1217$ &$1.1617$ &$1.2098$ &$1.2709$ & $1.6709$    \\
                                & 1000 &   $1.1599$ &$1.1439$ &$1.1148$ &$1.1116$ &$1.0928$& $1.1086$ &$1.1272$& $1.6004$   \\ 
                                 & 1500 &   $1.1547$& $1.1196$ &$1.1027$& $1.0757$ &$1.0625$ &$1.0587$& $1.1004$& $1.5725$  \\ 
                                  & 2000 &   $1.2721$ &$1.2310$& $1.1693$& $1.1195$& $1.1050$ &$1.1127$ &$1.1480$ &$1.5855$   \\                      \hline
       \multirow{4}{*}{\textbf{Scenario 3}}  & 500 &  $1.3609$ &$1.3427$& $1.3148$ &$1.2875$ &$1.2733$ &$1.3452$ &$1.4008$ & $1.6202$    \\
                                & 1000 &   $1.2853$ &$1.2603$ &$1.2287$ &$1.1773$ &$1.1989$& $1.2368$ &$1.2741$& $1.5454$   \\ 
                                 & 1500 &   $1.2490$& $1.2161$ &$1.2172$& $1.1716$ &$1.1862$ &$1.2017$& $1.2092$& $1.5266$  \\ 
                                  & 2000 &   $1.3389$ &$1.2699$& $1.2194$& $1.1751$& $1.1485$ &$1.1763$ &$1.1737$ &$1.5183$   \\           
    \bottomrule
    \end{tabular}
\end{table}

\section{Real Data Examples}
\label{sec:realdata}

\subsection{Gene expression data analysis}
\label{subsec:gene}
In this section, we analyze gene expression levels across several tissues.
The analysis is based on the GTEx Portal dataset, which includes gene expression data from 11,688 postmortem samples, covering 53 tissues from 714 human donors. Following \cite{li2022transfer}, we investigate how genes that are associated with the central nervous system (CNS) influence the expression of the protein-coding gene \textit{JAM2} (Junctional Adhesion Molecule B) in each tissue. After preprocessing, which involves matching recorded CNS genes and excluding samples with missing values, we obtain $1646$ genes as covariates. The gene expression levels often share similar structures across tissues and have their own characteristics in different tissues \citep{mele2015human,kryuchkova2017benchmark}. Given sufficient data in brain tissues, we use all 13 brain tissues in the source models, comprising a total of $n_s = 3220$ samples. The remaining tissues serve as targets, with one tissue selected as the target at a time. The target sample size $n_t$ ranges from $77$ to $816$. Both the covariates and the response gene are normalized using the estimated means and variances from the source data. 

It is known that gene expression data are often confounded by cell-type composition, batch effects and other latent factors \citep{jew2020accurate,ye2023batch}. \red{To demonstrate that our method can effectively adjust for these confounders, we construct a deconfounded estimator using additional data as the gold standard.} The dataset includes several genotyping principal components and PEER factors that can serve as proxies for hidden confounders. We select $q=20$ genotyping principal components and PEER factors as proxies for confounders and deconfound the dataset by regressing them out. A standard transfer learning procedure for high-dimensional linear regression is then applied to estimate the coefficient $\beta_t$ for each target tissue.

\begin{table}[ht]
\renewcommand{\arraystretch}{0.65}
\footnotesize
    \centerfloat
    \caption{The $l_2$ norm errors for estimating $\beta_t$ using the proposed methods and baselines for different target tissues.}
    \label{tab:gene_estimation}
    \begin{tabular}{ccccccccc}
        \toprule
        \textbf{Tissue} & Adipose  &Artery & Blood &Heart &  Lung & Muscle & Spleen&  Testis\\
        \midrule
        SingleTrim &0.4921 & 0.1838 &0.2518 &0.1451 &  1.4388 & 0.1546 & 0.1364 & 0.2318 \\
        TransTrim &0.3837 & 0.1428 &0.2084 &0.1427 &  1.2061 & 0.1050 & 0.1230&   0.2262  \\
        SingleFarm &  0.4826 & 0.1807 &0.2439 &0.1543 &1.3647 & 0.1484& 0.1284 &    0.2017  \\
        TransFarm & 0.2337 & 0.1671 &0.2957 & 0.1108 & 1.1777 & 0.1072 & 0.1090&   0.2416 \\
        \vspace{-0.25cm}\makecell[c]{ProTrans \vspace{-0.5cm}\\ (full source)} &0.1859 & 0.1247 &$\bm{0.1900}$ &  $\bm{0.0874}$ & 1.2296 & $\bm{0.0936}$ & $\bm{0.0943}$&  0.1659  \\
        \makecell[c]{ProTrans \vspace{-0.5cm}\\ (selected source)} & $\bm{0.1393}$ & $\bm{0.1179}$ & 0.2023 & 0.1134 & $\bm{0.2641}$ & 0.1180 & 0.1098 & $\bm{0.0902}$\\
        \bottomrule
    \end{tabular}
\end{table}

ProTrans and the four baseline approaches are implemented to estimate $\beta_t$ for each target model. Since there are $K = 13$ source models, we also apply the source selection procedure in Section \ref{sec:source_selection} with $\rho =1.1$. This setting ensures a relatively large yet informative source dataset, selecting between 1 and 5 of the most similar sources for each target tissue. We evaluate the performance of ProTrans and baselines using the $l_2$ norm error between the estimated coefficient and the gold standard. Table \ref{tab:gene_estimation} presents the results for several selected target tissues. ProTrans, either with or without informative source selection, outperforms the baseline methods, indicating that it is minimally affected by hidden confounding in this analysis. \red{Due to unobserved confounders, baseline methods often produce spurious associations involving certain gene-expression patterns. For example, in the target tissue Adipose, all four baseline methods incorrectly identify genes such as \textit{IGFBP6}, \textit{ASCL1} and \textit{FGB} as being associated with \textit{JAM2} expression. In contrast, both ProTrans and the gold-standard analysis detect no significant differential influence of these genes on \textit{JAM2} expression between the target and source tissues. This result is also consistent with the known biological functions of these genes reported in previous studies.}

The informative source selection further enhances the robustness of ProTrans. For target tissues such as blood, heart, muscle and spleen, where most source data are informative, the selection step may reduce the total source sample size and perform slightly worse than the unselected version. However, when the source data are partially informative for target tissues such as adipose, lung and testis, the selection procedure effectively filters out uninformative data, leading to notable improvements.

\subsection{Treatment effect of education on earnings}
We estimate the conditional average treatment effect (CATE) of education on earnings using data from the National Longitudinal Survey (NLS) of Young Men. This dataset contains responses from individuals aged 14 to 24 in 1966, along with their educational attainment and wages recorded during the 1976 follow-up. For preprocessing, following previous studies \citep{wang2018bounded,sun2022high}, we define the treatment variable $T$ as whether an individual attained education beyond a four-year college degree ($T=1$, high education) or not ($T=0$, low education). We retain only samples with valid information on education, wages, and select $17$ covariates after a screening step. Missing values in the selected covariates are imputed using mean imputation. Covariates and the response variable are normalized using the estimated means and variances from the source data. The final dataset includes 1,335 control group samples ($T=0$) and 680 treatment group samples ($T=1$).

\red{To apply ProTrans, we treat the control group as the source model and the treatment group as the target. We implement ProTrans and the baseline method based on \cite{tang2023synthetic} under the fixed-dimensional setting described in the Supplementary Material to estimate the conditional treatment effect. Similar to Section \ref{subsec:gene}, we construct a gold standard for the conditional treatment effect using additional data as proxies for confounders, including age, IQ scores, and residence in the South (all measured in 1966) as recommended by \cite{wang2018bounded}. The baseline method, which is strongly influenced by confounding, exhibits various forms of spurious association, such as effects of job type and job characteristics (survey questions R0029900, R0035500, and R0062600). In contrast, both our method and the gold standard indicate that these covariates do not significantly contribute to differences in earnings between the treatment and control groups once confounding factors are removed. Moreover, after adjusting for confounders, the treatment consistently shows a positive effect on earnings across covariate profiles.
}

\vspace{-0.5cm}
\section*{Disclosure Statement}\label{disclosure-statement}

The authors report there are no competing interests to declare. 
\vspace{-0.5cm}

\section*{Data Availability Statement}\label{data-availability-statement}
The data that support the findings of this study are publicly available at \href{https://www.nlsinfo.org/investigator/pages/search?s=NLSM}{https://www.nlsinfo.} \href{https://www.nlsinfo.org/investigator/pages/search?s=NLSM}{org/investigator/pages/search?s=NLSM} (NLS of Young Men) and \href{https://gtexportal.org/home/downloads/adult-gtex/bulk_tissue_expression}{https://gtexportal.org/ho} \href{https://gtexportal.org/home/downloads/adult-gtex/bulk_tissue_expression}{me/downloads/adult-gtex/bulk\_tissue\_expression} (GTEx Portal dataset).

\begin{center}
{\large\bf SUPPLEMENTARY MATERIAL}
\end{center}

\begin{description}

\item[Deconfounding\_via\_Profiled\_Transfer\_Learning\_supp]  Theoretical results and additional numerical results for ``Deconfounding via Profiled Transfer Learning'', (PDF file).

\item[Code and Data] R code to implement and reproduce the simulation and real data results, corresponding outputs and raw datasets. 

\end{description}


\setlength{\baselineskip}{0.85\baselineskip}
\bibliographystyle{agsm}
\bibliography{paper-ref}

@article{cevid2020spectral,
  title={Spectral deconfounding via perturbed sparse linear models},
  author={{\'C}evid, Domagoj and B{\"u}hlmann, Peter and Meinshausen, Nicolai},
  journal={Journal of Machine Learning Research},
  volume={21},
  number={232},
  pages={1--41},
  year={2020}
}

@article{guo2022doubly,
  title={Doubly debiased lasso: High-dimensional inference under hidden confounding},
  author={Guo, Zijian and {\'C}evid, Domagoj and B{\"u}hlmann, Peter},
  journal={Annals of statistics},
  volume={50},
  number={3},
  pages={1320},
  year={2022},
  publisher={NIH Public Access}
}

@article{li2022transfer,
  title={Transfer learning for high-dimensional linear regression: Prediction, estimation and minimax optimality},
  author={Li, Sai and Cai, T Tony and Li, Hongzhe},
  journal={Journal of the Royal Statistical Society Series B: Statistical Methodology},
  volume={84},
  number={1},
  pages={149--173},
  year={2022},
  publisher={Oxford University Press}
}

@article{li2024estimation,
  title={Estimation and inference for high-dimensional generalized linear models with knowledge transfer},
  author={Li, Sai and Zhang, Linjun and Cai, T Tony and Li, Hongzhe},
  journal={Journal of the American Statistical Association},
  volume={119},
  number={546},
  pages={1274--1285},
  year={2024},
  publisher={Taylor \& Francis}
}

@article{wang2018bounded,
  title={Bounded, efficient and multiply robust estimation of average treatment effects using instrumental variables},
  author={Wang, Linbo and Tchetgen Tchetgen, Eric},
  journal={Journal of the Royal Statistical Society Series B: Statistical Methodology},
  volume={80},
  number={3},
  pages={531--550},
  year={2018},
  publisher={Oxford University Press}
}

@article{sun2022high,
  title={High-dimensional model-assisted inference for local average treatment effects with instrumental variables},
  author={Sun, Baoluo and Tan, Zhiqiang},
  journal={Journal of Business \& Economic Statistics},
  volume={40},
  number={4},
  pages={1732--1744},
  year={2022},
  publisher={Taylor \& Francis}
}

@article{ye2023batch,
  title={Batch-effect correction with sample remeasurement in highly confounded case-control studies},
  author={Ye, Hanxuan and Zhang, Xianyang and Wang, Chen and Goode, Ellen L and Chen, Jun},
  journal={Nature computational science},
  volume={3},
  number={8},
  pages={709--719},
  year={2023},
  publisher={Nature Publishing Group US New York}
}

@article{jew2020accurate,
  title={Accurate estimation of cell composition in bulk expression through robust integration of single-cell information},
  author={Jew, Brandon and Alvarez, Marcus and Rahmani, Elior and Miao, Zong and Ko, Arthur and Garske, Kristina M and Sul, Jae Hoon and Pietil{\"a}inen, Kirsi H and Pajukanta, P{\"a}ivi and Halperin, Eran},
  journal={Nature communications},
  volume={11},
  number={1},
  pages={1971},
  year={2020},
  publisher={Nature Publishing Group UK London}
}

@article{fan2021robust,
  title={Robust high dimensional factor models with applications to statistical machine learning},
  author={Fan, Jianqing and Wang, Kaizheng and Zhong, Yiqiao and Zhu, Ziwei},
  journal={Statistical science: a review journal of the Institute of Mathematical Statistics},
  volume={36},
  number={2},
  pages={303},
  year={2021}
}

@article{weiss2016survey,
  title={A survey of transfer learning},
  author={Weiss, Karl and Khoshgoftaar, Taghi M and Wang, DingDing},
  journal={Journal of Big data},
  volume={3},
  pages={1--40},
  year={2016},
  publisher={Springer}
}

@article{cai2024transferb,
  title={Transfer learning for functional mean estimation: Phase transition and adaptive algorithms},
  author={Cai, T Tony and Kim, Dongwoo and Pu, Hongming},
  journal={The Annals of Statistics},
  volume={52},
  number={2},
  pages={654--678},
  year={2024},
  publisher={Institute of Mathematical Statistics}
}

@article{he2024adatrans,
  title={AdaTrans: Feature-wise and Sample-wise Adaptive Transfer Learning for High-dimensional Regression},
  author={He, Zelin and Sun, Ying and Liu, Jingyuan and Li, Runze},
  journal={arXiv preprint:2403.13565},
  year={2024}
}

@article{gholizade2025review,
  title={A review of recent advances and strategies in transfer learning},
  author={Gholizade, Masoume and Soltanizadeh, Hadi and Rahmanimanesh, Mohammad and Sana, Shib Sankar},
  journal={International Journal of System Assurance Engineering and Management},
  pages={1--40},
  year={2025},
  publisher={Springer}
}

@article{baiocchi2014instrumental,
  title={Instrumental variable methods for causal inference},
  author={Baiocchi, Michael and Cheng, Jing and Small, Dylan S},
  journal={Statistics in medicine},
  volume={33},
  number={13},
  pages={2297--2340},
  year={2014},
  publisher={Wiley Online Library}
}

@article{martens2006instrumental,
  title={Instrumental variables: application and limitations},
  author={Martens, Edwin P and Pestman, Wiebe R and de Boer, Anthonius and Belitser, Svetlana V and Klungel, Olaf H},
  journal={Epidemiology},
  volume={17},
  number={3},
  pages={260--267},
  year={2006},
  publisher={LWW}
}

@article{maciejewski2019using,
  title={Using instrumental variables to address bias from unobserved confounders},
  author={Maciejewski, Matthew L and Brookhart, M Alan},
  journal={Jama},
  volume={321},
  number={21},
  pages={2124--2125},
  year={2019},
  publisher={American Medical Association}
}

@article{miao2018identifying,
  title={Identifying causal effects with proxy variables of an unmeasured confounder},
  author={Miao, Wang and Geng, Zhi and Tchetgen Tchetgen, Eric J},
  journal={Biometrika},
  volume={105},
  number={4},
  pages={987--993},
  year={2018},
  publisher={Oxford University Press}
}

@article{cui2024semiparametric,
  title={Semiparametric proximal causal inference},
  author={Cui, Yifan and Pu, Hongming and Shi, Xu and Miao, Wang and Tchetgen Tchetgen, Eric},
  journal={Journal of the American Statistical Association},
  volume={119},
  number={546},
  pages={1348--1359},
  year={2024},
  publisher={Taylor \& Francis}
}

@article{ding2019bracketing,
  title={A bracketing relationship between difference-in-differences and lagged-dependent-variable adjustment},
  author={Ding, Peng and Li, Fan},
  journal={Political Analysis},
  volume={27},
  number={4},
  pages={605--615},
  year={2019},
  publisher={Cambridge University Press}
}

@article{donald2007inference,
  title={Inference with difference-in-differences and other panel data},
  author={Donald, Stephen G and Lang, Kevin},
  journal={The review of Economics and Statistics},
  volume={89},
  number={2},
  pages={221--233},
  year={2007},
  publisher={The MIT Press}
}

@article{fox2002structural,
  title={Structural equation models},
  author={Fox, John and others},
  journal={Appendix to an R and S-PLUS Companion to Applied Regression},
  year={2002},
  publisher={Sage Publications California}
}

@article{pearl2012causal,
  title={The causal foundations of structural equation modeling},
  author={Pearl, Judea},
  journal={Handbook of structural equation modeling},
  pages={68--91},
  year={2012}
}

@article{tian2023transfer,
  title={Transfer learning under high-dimensional generalized linear models},
  author={Tian, Ye and Feng, Yang},
  journal={Journal of the American Statistical Association},
  volume={118},
  number={544},
  pages={2684--2697},
  year={2023},
  publisher={Taylor \& Francis}
}

@article{sun2024decorrelating,
  title={A decorrelating and debiasing approach to simultaneous inference for high-dimensional confounded models},
  author={Sun, Yinrui and Ma, Li and Xia, Yin},
  journal={Journal of the American Statistical Association},
  volume={119},
  number={548},
  pages={2857--2868},
  year={2024},
  publisher={Taylor \& Francis}
}

@article{bing2022adaptive,
  title={Adaptive estimation in multivariate response regression with hidden variables},
  author={Bing, Xin and Ning, Yang and Xu, Yaosheng},
  journal={The annals of statistics},
  volume={50},
  number={2},
  pages={640--672},
  year={2022},
  publisher={Institute of Mathematical Statistics}
}

@article{zhao2025semiparametric,
  title={A semiparametric instrumented difference-in-differences approach to policy learning},
  author={Zhao, Pan and Cui, Yifan},
  journal={Biometrika},
  pages={asaf043},
  year={2025},
  publisher={Oxford University Press}
}

@article{mele2015human,
  title={The human transcriptome across tissues and individuals},
  author={Mel{\'e}, Marta and Ferreira, Pedro G and Reverter, Ferran and DeLuca, David S and Monlong, Jean and Sammeth, Michael and Young, Taylor R and Goldmann, Jakob M and Pervouchine, Dmitri D and Sullivan, Timothy J and others},
  journal={Science},
  volume={348},
  number={6235},
  pages={660--665},
  year={2015},
  publisher={American Association for the Advancement of Science}
}

@article{kryuchkova2017benchmark,
  title={A benchmark of gene expression tissue-specificity metrics},
  author={Kryuchkova-Mostacci, Nadezda and Robinson-Rechavi, Marc},
  journal={Briefings in bioinformatics},
  volume={18},
  number={2},
  pages={205--214},
  year={2017},
  publisher={Oxford University Press}
}

@article{baranowski2020ranking,
  title={Ranking-based variable selection for high-dimensional data},
  author={Baranowski, Rafal and Chen, Yining and Fryzlewicz, Piotr},
  journal={Statistica Sinica},
  volume={30},
  number={3},
  pages={1485--1516},
  year={2020},
  publisher={JSTOR}
}

@article{malhotra2021threshold,
  title={Threshold benchmarking for feature ranking techniques},
  author={Malhotra, Ruchika and Sharma, Anjali},
  journal={Bulletin of Electrical Engineering and Informatics},
  volume={10},
  number={2},
  pages={1063--1070},
  year={2021}
}

@article{tang2023synthetic,
  title={The synthetic instrument: From sparse association to sparse causation},
  author={Tang, Dingke and Kong, Dehan and Wang, Linbo},
  journal={arXiv preprint arXiv:2304.01098},
  year={2023}
}

@article{chernozhukov2017lava,
  title={A lava attack on the recovery of sums of dense and sparse signals},
  author={Chernozhukov, Victor and Hansen, Christian and Liao, Yuan},
  journal={The Annals of Statistics},
  volume={45},
  number={1},
  pages={39--76},
  year={2017}
}

@article{lin2024profiled,
  title={Profiled transfer learning for high dimensional linear model},
  author={Lin, Ziqian and Zhao, Junlong and Wang, Fang and Wang, Hansheng},
  journal={arXiv preprint arXiv:2406.00701},
  year={2024}
}

@article{tian2025learning,
  title={Learning from similar linear representations: Adaptivity, minimaxity, and robustness},
  author={Tian, Ye and Gu, Yuqi and Feng, Yang},
  journal={Journal of Machine Learning Research},
  volume={26},
  number={187},
  pages={1--125},
  year={2025}
}

@article{yuan2025optimal,
  title={Optimal Transport based Cross-Domain Integration for Heterogeneous Data},
  author={Yuan, Yubai and Zhang, Yijiao and Shahbaba, Babak and Fortin, Norbert and Cooper, Keiland and Nie, Qing and Qu, Annie},
  journal={Journal of the American Statistical Association},
  volume={120},
  number={551},
  pages={1449--1462},
  year={2025},
  publisher={Taylor \& Francis}
}
\end{document}